\documentclass[aps,prb,twocolumn,superscriptaddress,showpacs,floatfix,reprint]{revtex4-1}

\usepackage{amsmath}
\usepackage{amssymb}
\usepackage{graphicx}

\newcommand{\sgn}{\mathop{\rm sgn}\nolimits}

\newcommand{\ga}{{\alpha}}
\newcommand{\gt}{{\tau}}
\newcommand{\gx}{{\chi}}

\newcommand{\gvf}{{\varphi}}
\newcommand{\go}{{\omega}}

\newcommand{\gk}{{\kappa}}

\newcommand{\ac}{{\rm ac}}
\newcommand{\dc}{{\rm dc}}

\newcommand{\cS}{{\cal  S}}

\begin{document}

\title{Control of electron-hole pair generation by\\
biharmonic voltage drive of a quantum point contact}

\author{Mihajlo Vanevi\' c}%
\affiliation{Department of Physics, University of
Belgrade, Studentski trg 12, 11158 Belgrade, Serbia}

\author{Wolfgang Belzig}
\affiliation{Fachbereich Physik, Universit\" at Konstanz, D-78457
Konstanz, Germany}

\date{\today}

\begin{abstract}
A time-dependent electromagnetic field creates electron-hole 
excitations in a Fermi sea at low temperature. We show that
the electron-hole pairs can be generated in a controlled way using
harmonic and biharmonic time-dependent voltages applied to a
quantum contact and obtain the probabilities of the pair
creations. For a biharmonic voltage drive, we find that the
probability of a pair creation decreases in the presence of an
in-phase second harmonic. This accounts for the suppression of the
excess noise observed experimentally [Gabelli and Reulet,
arXiv:1205.3638] proving that dynamic control and detection of
elementary excitations in quantum conductors are within the reach
of the present technology.
\end{abstract}




\maketitle


The controllable generation and manipulation of single to few-particle
excitations in mesoscopic conductors has attracted much attention
recently.%
\cite{hermelin_electrons_2011,mcneil_-demand_2011,%
FeveMaheBerroirKontosGlattliEtienneSCIENCE316-07,%
mahe_current_2010,bocquillon_electron_2012,%
parmentier_current_2012,%
sherkunov_optimal_2012,%
gabelli_shaping_2012,%
VanevicNazarovBelzigPRL99-07,*VanevicNazarovBelzigPRB78-08}
The interest in the subject stems from the emerging field of
electron quantum optics, which has the goal to
develop coherent electronic devices suitable for
quantum communication and quantum computation.
This is based on the analogy with photon quantum optics where pairs
of photons in entangled polarization state are used for
transmission and processing of quantum information. In a
solid-state system, a similar functionality can be achieved using
pairs of quasiparticles with entangled spin or orbital degrees of
freedom.%
\cite{beenakker_proposal_2003,samuelsson_two-particle_2004}
An important building block of electronic devices aimed at quantum
computation is a tunable electron source capable of creating
quasiparticles in a coherent way. An on-demand coherent electron
source has been realized
recently using a localized electronic
level in a quantum dot, weakly coupled to a conductor, which is
populated and emptied by a modulation of its energy
via a periodic gate
voltage.\cite{FeveMaheBerroirKontosGlattliEtienneSCIENCE316-07}
This results in a sequence of quantized single-electron current
pulses whose properties can be inferred from the measurements
of the current noise spectrum.%
\cite{parmentier_current_2012,KeelingShytovLevitovPRL101-08}

An altogether different route which does not require electron
confinement has been suggested theoretically by Keeling, Klich,
and Levitov.\cite{KeelingKlichLevitovPRL97-06}
The authors have proposed to make use of time-dependent voltage
pulses to create excitations from a degenerate
Fermi sea in a mesoscopic conductor. It has been found that
Lorentzian-shaped pulses $V(t)$ of a quantized area
$\int eV(t) dt/\hbar = 2\pi N$ ($N$ is integer, $e$ is the electron
charge, and $\hbar$ is the reduced Planck constant,
hereafter $\hbar=1$) create exactly $N$
electrons above the Fermi level with no additional excitations.
The many-body quantum state created by these pulses is a product
state of $N$ particles added to an unperturbed Fermi sea,
independent on the relative position of the pulses, their duration, or
overlap. Conversely, the charge transfer statistics in this case
is binomial, which supports the picture of independent charge
quanta created. The single-particle character of excitations can
be probed by noise measurements: it manifests itself as a
reduction of the current noise power which assumes its minimal
value set by the dc
voltage offset.\cite{IvanovLeeLevitovPRB56-97}
%
%
In general, however, a time-dependent field does create additional
excitations in a Fermi sea. These excitations give rise to excess
photon-assisted noise, exceeding the minimal dc
noise level.\cite{LesovikLevitovPRL72-94,PedersenButtikerPRB58-98}
The photon-assisted noise was observed experimentally in
diffusive phase-coherent metallic conductors,%
\cite{SchoelkopfKozhevnikovProberRooksPRL80-98}
normal-metal--superconductor junctions,%
\cite{KozhevnikovProberPRL84-99}
and quantum point contacts\cite{ReydelletRocheGlattliEtienneJinPRL90-03}
with harmonic ac-voltage applied.
Noise spectroscopy of a quantum tunnel junction
with a more complex biharmonic voltage drive
has been carried out recently.\cite{gabelli_shaping_2012}

The physical picture behind photon-assisted noise has been
revealed in Ref.%
~\onlinecite{VanevicNazarovBelzigPRL99-07,*VanevicNazarovBelzigPRB78-08}
by an analysis of the full counting statistics: The elementary
excitations in the system at low temperature are the
{\em electron-hole pairs} created by the ac voltage component,
in addition to electrons injected by the dc voltage offset. This
picture is valid beyond noise measurements and pertains to the
full statistics of the transferred charge.
Since electrons and holes from a pair are created with the same
probability, the pairs give rise only to the excess noise and
higher even-order current correlators, whereas they give no
contribution to the average current and odd-order correlators.
The number of created electron-hole pairs depends on
the shape and the amplitude of the ac voltage applied.
This opens a route toward dynamic control of elementary
excitations in quantum conductors which, if proved feasible,
could be used in the electron-hole sources to produce quasiparticles
with entangled spin or orbital degrees of freedom.%
\cite{sherkunov_optimal_2012,%
RychkovPolianskiButtikerPRB72-05,%
PolianskiSamuelssonBuettikerPRB72-05,samuelsson_dynamic_2005,%
beenakker_optimal_2005,LebedevLesovikBlatterPRB72-05,%
splettstoesser_two-particle_2009,moskalets_spectroscopy_2011,%
grenier_single-electron_2011}
Alternatively, such control can also be used to minimize
excitations present in the system and approach the limit of an
ideal single-electron source using realistic voltage pulses.

In this paper, we investigate the feasibility of a dynamic control
of elementary electron-hole pair excitations.
We analyze experimental data on the photon-assisted noise in
quantum conductors subject to time dependent drive,%
\cite{ReydelletRocheGlattliEtienneJinPRL90-03,gabelli_shaping_2012}
identify elementary excitations generated in the experiments,
and show how the measured excess noise is composed of the
contributions of electron-hole pairs created. For a quantum
contact with harmonic voltage drive studied in
Ref.~\onlinecite{ReydelletRocheGlattliEtienneJinPRL90-03},
we find that a single electron-hole pair is created per period
with a certain probability, in addition to the electrons that are
injected by dc voltage offset. We find how the probability of pair
creation depends on the amplitude of the ac voltage drive, which
is in agreement with the observed excess noise.

In the recent experiment by Gabelli and
Reulet,\cite{gabelli_shaping_2012}
the authors have studied photon-assisted noise in a quantum tunnel
junction subject to a biharmonic time-dependent voltage, where the
dc offset, the ac amplitudes, and the relative phase are tunable.
The authors have observed the
reduction of the noise when an in-phase second harmonic is present
in the drive. We find that the statistics of the transferred
charge is the simplest when the dc offset is an integer multiple
of the driving frequency. In that case only a few (one or two)
electron-hole pairs are created per period with probabilities that
depend on the shape and the amplitude of the biharmonic voltage
component. In addition,  we relate the reduction of the noise in
the presence of an in-phase second harmonic to the suppressed
probability of the electron-hole pair creation as the drive
voltage approaches the shape of optimal Lorentzian pulses. The
agreement with the observed excess noise corroborates that the
dynamic control of elementary excitations in quantum conductors
has been achieved experimentally.


The system we study is a coherent mesoscopic conductor
characterized by transmission eigenvalues $\{ T_n \}$
where $n$ labels spin-degenerate transport channels. Conductance
of the conductor is $G=(e^2/\pi)\sum_n T_n$ and the
Fano factor $F=(\sum_n T_n R_n)/\sum_n T_n$ where $R_n=1-T_n$ are
reflection probabilities. The conductor is subject to a
periodic voltage drive $V(t) = \bar V + V_\ac(t)$ where
$\bar V$ is dc voltage offset and $V_\ac(t)$ is the ac
voltage component with zero average and period $\gt = 2\pi/\go$.
%
\begin{figure}[t]
\includegraphics[scale=0.9]{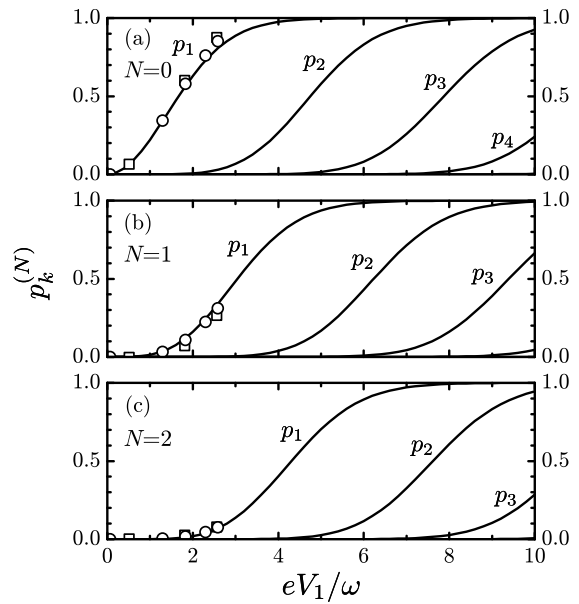}
\caption{\label{fig:pkV1a.eps}
    Probabilities $p_k$ of electron-hole pair creations for
    harmonic drive $V(t) = \bar V + V_1 \cos(\go t)$ as a function
    of the ac amplitude $V_1$ and for different dc voltage
    offsets $e\bar V/\go = N$:
    (a) $N=0$, (b) $N=1$, and (c) $N=2$.
    DC offset $\bar V$ sets the number of electrons injected
    towards the contact while $V_1$ controls the electron-hole pair
    generation.
    Symbols denote experimental data for the excess noise
    $S_\ac$ in units $2GF\go$ taken from
    Ref.~\onlinecite{ReydelletRocheGlattliEtienneJinPRL90-03}
    for a quantum contact with $F=1/2$ and
    ({\scriptsize $\bigcirc$}) $\go/2\pi = 17.32\, \rm GHz$,
    ($\square$) $\go/2\pi = 8.73\, \rm GHz$.
}
\end{figure}
%
At low temperature $T\ll \go$, the cumulant generating function
$\cS(\gx)$ of the charge transfer statistics can be cast in a
form that manifestly reveals what are the elementary processes
in the system:\cite{VanevicNazarovBelzigPRL99-07}
$\cS(\gx) = \cS_\dc(\gx) + \cS_\ac(\gx)$, where
$\cS_\dc = (t_0 |e\bar V|/\pi)
\sum_n \ln[ 1 + T_n(e^{-i\gk\gx}-1) ]$
and
\begin{align}
\cS_\ac
=&
\frac{2 t_0}{\gt} \sum_{n,k}
\left( (1-\bar v)
\ln[ 1 + p_k^{(N)} T_n R_n (e^{i\gx} + e^{-i\gx}-2)]
\right.  \notag
\\
&+
\left.
\bar v \, \ln[ 1 + p_k^{(N+1)} T_n R_n (e^{i\gx} + e^{-i\gx}-2)]
\right).
\end{align}
Here, $t_0$ is the measurement time which is much larger than the
characteristic time scale on which current fluctuations are
correlated,
$\gk = 1$ ($\gk = -1$) for $e\bar V>0$ ($e\bar V<0$)
is related to direction of the charge transfer,
and $N = \lfloor e\bar V/\go \rfloor$
($\bar v = e\bar V/\go - N$) is the integer (fractional) part
of $e\bar V/\go$. Coefficients
$p_k^{(N)}$ ($0\le p_k^{(N)} \le 1$) in $\cS_\ac$ are the
probabilities of electron-hole pair creations
which depend on the details of the time-dependent voltage applied,
see Fig.~\ref{fig:pkV1a.eps}.
They are given by $p_k^{(N)} = \sin^2(\ga_k/2)$,
where $e^{\pm i \ga_k}$ are the {\em pairs} of complex conjugate
eigenvalues of the matrix
$M_{nm} = \sgn(n+0^+) \sum_{k=-\infty}^\infty
a_{n+k} a^*_{m+k} \sgn(0^+-k-N)$
(cf. Ref.~\onlinecite{VanevicNazarovBelzigPRL99-07,VanevicNazarovBelzigPRB78-08}
for details). The Fourier coefficients $a_n=\gt^{-1}\int_0^\gt
dt\, e^{-i\phi(t)}e^{in\go t}$ characterize the drive, where
$\phi(t)=\int_0^t dt'\, eV_\ac(t')$ is the phase acquired.

\begin{figure}[t]
\includegraphics[scale=0.8]{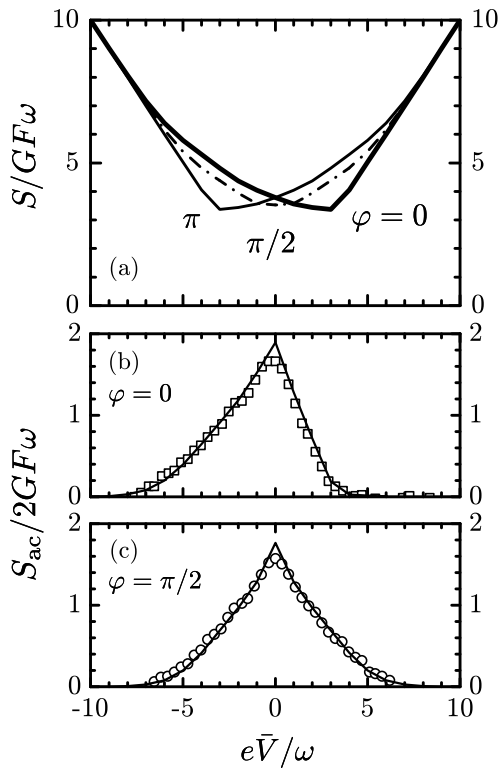}
\caption{\label{fig:SIvsVb}
    (a) Current noise $S$ as a function of dc voltage $\bar V$ for
    biharmonic time-dependent drive $V(t) = \bar V + V_1 \cos(\go
    t) + V_2 \cos(2\go t + \gvf)$ with $eV_1/\go = 5.4$, $eV_2/\go
    = 2.7$, and $\gvf = 0$ (thick solid line), $\pi/2$
    (dash-dotted line), and $\pi$ (thin solid line). The excess
    noise $S_\ac$ is shown as a function of $\bar V$ for the same
    voltage drive with (b) $\gvf = 0$ and (c) $\gvf = \pi/2$.
    Symbols in (b) and (c) denote experimental data taken from
    Ref.~\onlinecite{gabelli_shaping_2012} for a tunnel
    junction ($F=1$) at low temperature ($T=0.14\go$).
}
\end{figure}

\begin{figure}[t]
\includegraphics[scale=0.9]{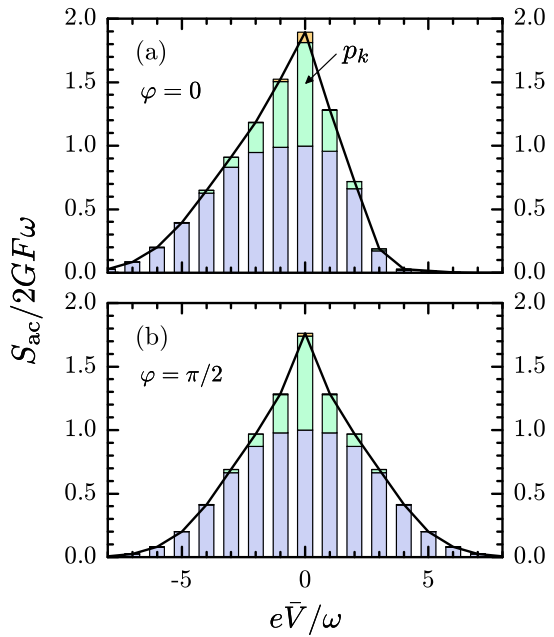}
\caption{\label{fig:pkPhi0_Pid2a}
    Decomposition of the excess noise $S_\ac$ shown in
    Fig.~\ref{fig:SIvsVb} into elementary electron-hole pair
    excitations [see Eq.~\eqref{eq:Sac-decomp-pk}].
    Probabilities $p_k^{(N)}$ of the pair creations are denoted by
    stacked bars at dc voltages $e\bar V/\go = N$ with $N$
    integer: $k=1$ (blue), $k=2$ (green), and $k=3$ (red bars).
    The elementary excitations comprise a few (one or two)
    electron-hole pairs generated with probabilities
    $p_{1,2}^{(N)}$  per voltage cycle.
}
\end{figure}

The physical interpretation of $\cS(\gx)$ is as follows.
The dc part $\cS_\dc(\gx)$ describes unidirectional single
electron transfers due to a finite voltage offset $\bar V$.
The attempt frequency is $|e\bar V|/\pi$ and the transfer
probability $T_n$ per transport channel.
The ac part of the cumulant generating function $\cS_\ac(\gx)$
describes the events of electron-hole pair creations.
The electron-hole pairs (labelled by $k$) are created by
the ac component of the drive with probability $p_k$ per voltage cycle.
The charge transfer occurs with probability $p_kT_nR_n$
when one particle -- e.g., electron -- is transmitted and the
hole is reflected, or vice versa. Since the electron and hole
from a pair are transmitted with equal probabilities,
the electron-hole pairs give no contribution to the average
current but they do contribute to the current noise power.
We note that the total number of attempts to create an
electron-hole pair is $t_0/\tau$ (per spin-split transport
channel), which gives one attempt per voltage cycle. These
attempts are shared
between processes $p_k^{(N)}$ and $p_k^{(N+1)}$ that correspond
to the nearest integer values of $e\bar V/\go$. The simplest
statistics -- determined by one type of the processes $p_k^{(N)}$
only -- is obtained for the voltage offset $e\bar V = N\go$
which is an integer multiple of the driving frequency.
For a nonzero dc component, the created electron-hole pairs
coexist with $N$ electrons injected towards the contact per period.

From the cumulant generating function we obtain
the current noise power $S=(e^2/t_0)\partial^2_{i\gx}\cS|_{\gx=0}
=S_\dc+S_\ac$ with $S_\dc = GF|e\bar V|$
and
$ S_\ac = 2GF\go \sum_k
\left( (1-\bar v)p_k^{(N)} + \bar v p_k^{(N+1)} \right)$.
Therefore, the excess noise of the electron-hole pairs
is a piecewise linear function of a dc voltage offset.
At $e\bar V/ \go = N$ ($N$ is integer), $S_\ac$
is given by the sum of the probabilities of the pair creations
\begin{equation}\label{eq:Sac-decomp-pk}
S_\ac|_N = 2GF\go \sum_k p_k^{(N)}
\end{equation}
and linearly interpolates in-between for $e\bar V/\go$ noninteger.
The experimentally observed excess noise%
\cite{ReydelletRocheGlattliEtienneJinPRL90-03,gabelli_shaping_2012}
and its decomposition into contributions of the electron-hole
pairs is shown in Figs.~\ref{fig:pkV1a.eps}~--~\ref{fig:pkPhi0_Pid2a}.


In Ref.~\onlinecite{ReydelletRocheGlattliEtienneJinPRL90-03}, the
authors measured photon-assisted noise in a quantum
contact ($F=1/2$) subject to harmonic time-dependent drive
$V(t) =\bar V + V_1 \cos(\go t)$. For this drive
$a_n = J_n(eV_1/\go)$ where $J_n(x)$ are Bessel functions of the
first kind. The predicted probabilities $p_k^{(N)}$ of the
electron-hole pair creations are shown in Fig.~\ref{fig:pkV1a.eps}
as a function of the ac amplitude $V_1$ and for different values
$e\bar V/\go =N$ of the dc component. As the drive amplitude $V_1$
is increased, the probability of the pair creation also increases
and more pairs are created per period. The symbols in
Fig.~\ref{fig:pkV1a.eps} denote experimental data for the
excess noise $S_\ac/2GF\go$ at dc offsets
$e\bar V/\go = N$ ($N=0,1,2$), measured for
driving frequencies $\go/2\pi = 17.32\,$GHz and $8.73\,$GHz.
The experimental results are in agreement with
Eq.~\eqref{eq:Sac-decomp-pk}. We find that for the amplitudes
used in the experiment only a single electron-hole pair is
created per period with probability $p_1$. The interval
$eV_1/\go$ where a single pair is excited is broad and is not
restricted to the weak driving limit $eV_1/\go\ll 1$, see
Fig.~\ref{fig:pkV1a.eps}~(a).
For a nonzero dc offset, the pair creation is also
accompanied by the single-electron transfers.
When the ac amplitude is smaller than the dc offset,
the creation of electron-hole pairs is strongly suppressed, see
Fig.~\ref{fig:pkV1a.eps}~(b),(c).
In the situation in which the probability of a pair creation is
small, the noise is dominated by the dc shot noise
$S_\dc \gg S_\ac$. In that case, generated electron-hole pairs can
be probed more directly in a beam splitter geometry where current
{\it cross correlation} occurs due to processes in which the pairs
are split and the two particles enter different outgoing terminals.%
\cite{VanevicNazarovBelzigPRB78-08,RychkovPolianskiButtikerPRB72-05}

In a recent experiment, Gabelli and
Reulet\cite{gabelli_shaping_2012} studied the noise in a
quantum tunnel junction ($F=1$) with a biharmonic voltage applied,
$V(t) = \bar V + V_1 \cos(\go t) + V_2 \cos(2\go t + \gvf)$,
where the dc offset $\bar V$, the ac amplitudes $V_1$ and $V_2$,
and the phase shift $\gvf$ were tunable.
The coefficients $a_n$ in this case are given by
$a_n=\sum_{m=-\infty}^\infty
J_{n-2m}(eV_1/\go)J_m(eV_2/2\go)e^{-im\gvf}$.
The observed current noise power and the decomposition of the
excess noise into contributions of the electron-hole pairs are
shown in Figs.~\ref{fig:SIvsVb} and~\ref{fig:pkPhi0_Pid2a}.
%
\begin{figure}[t]
\includegraphics[scale=0.9]{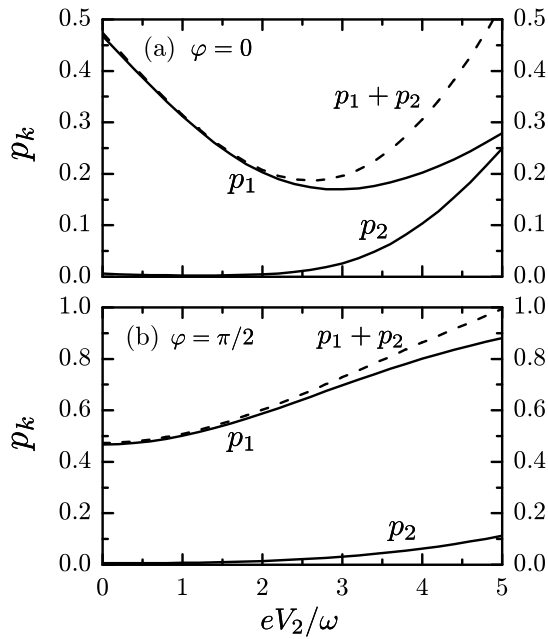}
\caption{\label{fig:pkV2Phi0_Pid2}
    Probabilities $p_k$ of the electron-hole pair creations as a
    function of the amplitude $V_2$ of the second harmonic for the
    biharmonic voltage drive with $e\bar V/\go = 3$, $eV_1/\go =
    5.4$, and the phase difference (a) $\gvf=0$ and (b) $\gvf =
    \pi/2$. Dashed lines indicate the excess noise $S_\ac$ in
    units $2GF\go$. The presence of an in-phase second harmonic
    decreases the probability of the electron-hole pair creation
    leading to suppression of $S_\ac$ observed
    experimentally.\cite{gabelli_shaping_2012}
}
\end{figure}
%
The current noise power as a function of $\bar V$ and for different
phase shifts $\gvf$ is shown in Fig.~\ref{fig:SIvsVb} (a),
where we use the experimental values of the parameters
$eV_1=5.4\go$ and $eV_2=2.7\go$. At low temperature $T\ll \go$,
the current noise power reads\cite{LesovikLevitovPRL72-94}
$S=GF\sum_{n=-\infty}^\infty |e\bar V + n\go|\, |a_n|^2$,
which is a piecewise linear function of the dc voltage offset
$e\bar V$ with kinks at integer multiples of $\go$.
This can be seen in the differential noise
$S'=\partial S/\partial(e\bar V)$ which consists of a series
of steps at $e\bar V/\go = N$ ($N$ is integer)
with the step size 
$S'|_{N+0}-S'|_{N-0} = 2GF|a_{-N}|^2$ (cf. Fig.~3 in
Ref.~\onlinecite{gabelli_shaping_2012}).
The excess noise is $S_\ac|_N = 2GF\go \sum_{n=1}^\infty n
|a_{-N\mp n}|^2$ where the upper (lower) sign is taken for
$N\ge 0$ ($N<0$). This analysis quantitatively accounts
for the photon-assisted noise observed experimentally,
see Fig.~\ref{fig:SIvsVb} (b), (c).
However, the information on elementary excitations present in the
system is implicit and encoded in the Fourier components $a_n$
characterizing the drive.

The decomposition of $S_\ac$ into contributions of electron-hole
pair excitations given by Eq.~\eqref{eq:Sac-decomp-pk} is shown in
Fig.~\ref{fig:pkPhi0_Pid2a}. The probabilities $p_k^{(N)}$ of the
pair creations are depicted as stacked bars. We observe that the
excess noise consists of a small number (one or two) of the
electron-hole pairs excited per period with certain probabilities.
The pairs are accompanied with $e\bar V/\go = N$ electrons
injected towards the contact per period. Therefore, for
small offset voltages the number of electron-hole pairs is
significant as compared to the number of electrons injected per
period. As the dc offset is increased, the probability of pair
creation decreases and the excess electrons start to dominate.
The probability of pair creation can further be tuned by
changing the shape of the ac voltage component.

In Ref.~\onlinecite{gabelli_shaping_2012}, the authors have used
a setup where the dc offset and the amplitude of the first harmonic
are kept fixed and the amplitude of the second harmonic $V_2$ is
varied to minimize the noise. In Fig.~\ref{fig:pkV2Phi0_Pid2} we
show the probabilities of the electron-hole pair generation as a
function of $V_2$ for a dc offset $e\bar V/\go = 3$
and the amplitude of the first harmonic $e V_1/\go = 5.4$.
For a simple harmonic drive ($V_2=0$) there is only one
electron-hole pair generated per period with probability $p_1$.
When the second harmonic is in phase with the first one
($\gvf=0$), the probability $p_1$ of the pair generation {\em
decreases} as the amplitude $V_2$ is increased,
Fig.~\ref{fig:pkV2Phi0_Pid2}~(a). As $V_2$ is increased further,
the second electron-hole pair is generated with the
increasing probability $p_2$ per period. For the amplitude
$eV_2/\go \approx 2.6$, the total
probability $p_1+p_2$ of the electron-hole pair creation
exhibits a minimum. This leads to the minimal excess noise in
Eq.~\eqref{eq:Sac-decomp-pk} which has been observed in
Ref.~\onlinecite{gabelli_shaping_2012}.
On the other hand, when the phase difference between the first and
the second harmonic is $\gvf=\pi/2$, the shape of the drive
deviates from the optimal one as $V_2$ is increased. The
probability of the electron-hole pair generation grows
monotonically with $V_2$ [see Fig.~\ref{fig:pkV2Phi0_Pid2}~(b)]
and the excess noise increases.\cite{gabelli_shaping_2012}


In conclusion, we have studied the available experimental data on
quantum noise in mesoscopic conductors with applied time-dependent
voltages and provided an interpretation
in terms of 
independent electrons and the electron-hole pairs created by the
drive. This interpretation is valid for a generic quantum
junction and to all orders in the statistics of the transferred
charge. We have shown how the excess photon-assisted noise is
composed of the contributions of the electron-hole pairs, whose
number and probability of creation can be controlled by changing
the shape and amplitude of the applied voltage. The agreement
of the predicted pair creation probabilities and the observed
photon-assisted noise corroborates that the dynamic control
of elementary excitations 
has been achieved experimentally. This can be utilized in
electron-hole sources that emit quasiparticles with entangled spin
or orbital degrees of freedom for use in mesoscopic electronics
and electron quantum optics.


We are grateful to Julien Gabelli, Yuli V. Nazarov, and Bertrand
Reulet for fruitful discussions. The research was supported by the
Serbian Ministry of Science, Project No.~171027, and the DFG
through SFB 767 and SPP1285.


\bibliography{FCSquantumControl}

\begin{thebibliography}{28}%
\makeatletter
\providecommand \@ifxundefined [1]{%
 \@ifx{#1\undefined}
}%
\providecommand \@ifnum [1]{%
 \ifnum #1\expandafter \@firstoftwo
 \else \expandafter \@secondoftwo
 \fi
}%
\providecommand \@ifx [1]{%
 \ifx #1\expandafter \@firstoftwo
 \else \expandafter \@secondoftwo
 \fi
}%
\providecommand \natexlab [1]{#1}%
\providecommand \enquote  [1]{``#1''}%
\providecommand \bibnamefont  [1]{#1}%
\providecommand \bibfnamefont [1]{#1}%
\providecommand \citenamefont [1]{#1}%
\providecommand \href@noop [0]{\@secondoftwo}%
\providecommand \href [0]{\begingroup \@sanitize@url \@href}%
\providecommand \@href[1]{\@@startlink{#1}\@@href}%
\providecommand \@@href[1]{\endgroup#1\@@endlink}%
\providecommand \@sanitize@url [0]{\catcode `\\12\catcode `\$12\catcode
  `\&12\catcode `\#12\catcode `\^12\catcode `\_12\catcode `\%12\relax}%
\providecommand \@@startlink[1]{}%
\providecommand \@@endlink[0]{}%
\providecommand \url  [0]{\begingroup\@sanitize@url \@url }%
\providecommand \@url [1]{\endgroup\@href {#1}{\urlprefix }}%
\providecommand \urlprefix  [0]{URL }%
\providecommand \Eprint [0]{\href }%
\providecommand \doibase [0]{http://dx.doi.org/}%
\providecommand \selectlanguage [0]{\@gobble}%
\providecommand \bibinfo  [0]{\@secondoftwo}%
\providecommand \bibfield  [0]{\@secondoftwo}%
\providecommand \translation [1]{[#1]}%
\providecommand \BibitemOpen [0]{}%
\providecommand \bibitemStop [0]{}%
\providecommand \bibitemNoStop [0]{.\EOS\space}%
\providecommand \EOS [0]{\spacefactor3000\relax}%
\providecommand \BibitemShut  [1]{\csname bibitem#1\endcsname}%
\let\auto@bib@innerbib\@empty
\bibitem [{\citenamefont {Hermelin}\ \emph {et~al.}(2011)\citenamefont
  {Hermelin}, \citenamefont {Takada}, \citenamefont {Yamamoto}, \citenamefont
  {Tarucha}, \citenamefont {Wieck}, \citenamefont {Saminadayar}, \citenamefont
  {B{\"a}uerle},\ and\ \citenamefont {Meunier}}]{hermelin_electrons_2011}%
  \BibitemOpen
  \bibfield  {author} {\bibinfo {author} {\bibfnamefont {S.}~\bibnamefont
  {Hermelin}}, \bibinfo {author} {\bibfnamefont {S.}~\bibnamefont {Takada}},
  \bibinfo {author} {\bibfnamefont {M.}~\bibnamefont {Yamamoto}}, \bibinfo
  {author} {\bibfnamefont {S.}~\bibnamefont {Tarucha}}, \bibinfo {author}
  {\bibfnamefont {A.~D.}\ \bibnamefont {Wieck}}, \bibinfo {author}
  {\bibfnamefont {L.}~\bibnamefont {Saminadayar}}, \bibinfo {author}
  {\bibfnamefont {C.}~\bibnamefont {B{\"a}uerle}}, \ and\ \bibinfo {author}
  {\bibfnamefont {T.}~\bibnamefont {Meunier}},\ }\href@noop {} {\bibfield
  {journal} {\bibinfo  {journal} {Nature}\ }\textbf {\bibinfo {volume} {477}},\
  \bibinfo {pages} {435} (\bibinfo {year} {2011})}\BibitemShut {NoStop}%
\bibitem [{\citenamefont {{McNeil}}\ \emph {et~al.}(2011)\citenamefont
  {{McNeil}}, \citenamefont {Kataoka}, \citenamefont {Ford}, \citenamefont
  {Barnes}, \citenamefont {Anderson}, \citenamefont {Jones}, \citenamefont
  {Farrer},\ and\ \citenamefont {Ritchie}}]{mcneil_-demand_2011}%
  \BibitemOpen
  \bibfield  {author} {\bibinfo {author} {\bibfnamefont {R.~P.}\ \bibnamefont
  {{McNeil}}}, \bibinfo {author} {\bibfnamefont {M.}~\bibnamefont {Kataoka}},
  \bibinfo {author} {\bibfnamefont {C.~J.~B.}\ \bibnamefont {Ford}}, \bibinfo
  {author} {\bibfnamefont {C.~H.~W.}\ \bibnamefont {Barnes}}, \bibinfo {author}
  {\bibfnamefont {D.}~\bibnamefont {Anderson}}, \bibinfo {author}
  {\bibfnamefont {G.~a.~C.}\ \bibnamefont {Jones}}, \bibinfo {author}
  {\bibfnamefont {I.}~\bibnamefont {Farrer}}, \ and\ \bibinfo {author}
  {\bibfnamefont {D.~A.}\ \bibnamefont {Ritchie}},\ }\href@noop {} {\bibfield
  {journal} {\bibinfo  {journal} {Nature}\ }\textbf {\bibinfo {volume} {477}},\
  \bibinfo {pages} {439} (\bibinfo {year} {2011})}\BibitemShut {NoStop}%
\bibitem [{\citenamefont {F{\`e}ve}\ \emph {et~al.}(2007)\citenamefont
  {F{\`e}ve}, \citenamefont {Mah{\'e}}, \citenamefont {Berroir}, \citenamefont
  {Kontos}, \citenamefont {Pla{\c c}ais}, \citenamefont {Glattli},
  \citenamefont {Cavanna}, \citenamefont {Etienne},\ and\ \citenamefont
  {Jin}}]{FeveMaheBerroirKontosGlattliEtienneSCIENCE316-07}%
  \BibitemOpen
  \bibfield  {author} {\bibinfo {author} {\bibfnamefont {G.}~\bibnamefont
  {F{\`e}ve}}, \bibinfo {author} {\bibfnamefont {A.}~\bibnamefont {Mah{\'e}}},
  \bibinfo {author} {\bibfnamefont {J.-M.}\ \bibnamefont {Berroir}}, \bibinfo
  {author} {\bibfnamefont {T.}~\bibnamefont {Kontos}}, \bibinfo {author}
  {\bibfnamefont {B.}~\bibnamefont {Pla{\c c}ais}}, \bibinfo {author}
  {\bibfnamefont {D.~C.}\ \bibnamefont {Glattli}}, \bibinfo {author}
  {\bibfnamefont {A.}~\bibnamefont {Cavanna}}, \bibinfo {author} {\bibfnamefont
  {B.}~\bibnamefont {Etienne}}, \ and\ \bibinfo {author} {\bibfnamefont
  {Y.}~\bibnamefont {Jin}},\ }\href@noop {} {\bibfield  {journal} {\bibinfo
  {journal} {Science}\ }\textbf {\bibinfo {volume} {316}},\ \bibinfo {pages}
  {1169} (\bibinfo {year} {2007})}\BibitemShut {NoStop}%
\bibitem [{\citenamefont {Mah{\'e}}\ \emph {et~al.}(2010)\citenamefont
  {Mah{\'e}}, \citenamefont {Parmentier}, \citenamefont {Bocquillon},
  \citenamefont {Berroir}, \citenamefont {Glattli}, \citenamefont {Kontos},
  \citenamefont {Pla{\c{c}}ais}, \citenamefont {F{\`e}ve}, \citenamefont
  {Cavanna},\ and\ \citenamefont {Jin}}]{mahe_current_2010}%
  \BibitemOpen
  \bibfield  {author} {\bibinfo {author} {\bibfnamefont {A.}~\bibnamefont
  {Mah{\'e}}}, \bibinfo {author} {\bibfnamefont {F.~D.}\ \bibnamefont
  {Parmentier}}, \bibinfo {author} {\bibfnamefont {E.}~\bibnamefont
  {Bocquillon}}, \bibinfo {author} {\bibfnamefont {J.}~\bibnamefont {Berroir}},
  \bibinfo {author} {\bibfnamefont {D.~C.}\ \bibnamefont {Glattli}}, \bibinfo
  {author} {\bibfnamefont {T.}~\bibnamefont {Kontos}}, \bibinfo {author}
  {\bibfnamefont {B.}~\bibnamefont {Pla{\c{c}}ais}}, \bibinfo {author}
  {\bibfnamefont {G.}~\bibnamefont {F{\`e}ve}}, \bibinfo {author}
  {\bibfnamefont {A.}~\bibnamefont {Cavanna}}, \ and\ \bibinfo {author}
  {\bibfnamefont {Y.}~\bibnamefont {Jin}},\ }\href@noop {} {\bibfield
  {journal} {\bibinfo  {journal} {Phys. Rev. B}\ }\textbf {\bibinfo {volume}
  {82}},\ \bibinfo {pages} {201309} (\bibinfo {year} {2010})}\BibitemShut
  {NoStop}%
\bibitem [{\citenamefont {Bocquillon}\ \emph {et~al.}(2012)\citenamefont
  {Bocquillon}, \citenamefont {Parmentier}, \citenamefont {Grenier},
  \citenamefont {Berroir}, \citenamefont {Degiovanni}, \citenamefont {Glattli},
  \citenamefont {Pla{\c{c}}ais}, \citenamefont {Cavanna}, \citenamefont {Jin},\
  and\ \citenamefont {F{\`e}ve}}]{bocquillon_electron_2012}%
  \BibitemOpen
  \bibfield  {author} {\bibinfo {author} {\bibfnamefont {E.}~\bibnamefont
  {Bocquillon}}, \bibinfo {author} {\bibfnamefont {F.~D.}\ \bibnamefont
  {Parmentier}}, \bibinfo {author} {\bibfnamefont {C.}~\bibnamefont {Grenier}},
  \bibinfo {author} {\bibfnamefont {J.}~\bibnamefont {Berroir}}, \bibinfo
  {author} {\bibfnamefont {P.}~\bibnamefont {Degiovanni}}, \bibinfo {author}
  {\bibfnamefont {D.~C.}\ \bibnamefont {Glattli}}, \bibinfo {author}
  {\bibfnamefont {B.}~\bibnamefont {Pla{\c{c}}ais}}, \bibinfo {author}
  {\bibfnamefont {A.}~\bibnamefont {Cavanna}}, \bibinfo {author} {\bibfnamefont
  {Y.}~\bibnamefont {Jin}}, \ and\ \bibinfo {author} {\bibfnamefont
  {G.}~\bibnamefont {F{\`e}ve}},\ }\href@noop {} {\bibfield  {journal}
  {\bibinfo  {journal} {Phys. Rev. Lett.}\ }\textbf {\bibinfo {volume} {108}},\
  \bibinfo {pages} {196803} (\bibinfo {year} {2012})}\BibitemShut {NoStop}%
\bibitem [{\citenamefont {Parmentier}\ \emph {et~al.}(2012)\citenamefont
  {Parmentier}, \citenamefont {Bocquillon}, \citenamefont {Berroir},
  \citenamefont {Glattli}, \citenamefont {Pla{\c{c}}ais}, \citenamefont {F{\`
  e}ve}, \citenamefont {Albert}, \citenamefont {Flindt},\ and\ \citenamefont
  {B{\"u}ttiker}}]{parmentier_current_2012}%
  \BibitemOpen
  \bibfield  {author} {\bibinfo {author} {\bibfnamefont {F.~D.}\ \bibnamefont
  {Parmentier}}, \bibinfo {author} {\bibfnamefont {E.}~\bibnamefont
  {Bocquillon}}, \bibinfo {author} {\bibfnamefont {J.}~\bibnamefont {Berroir}},
  \bibinfo {author} {\bibfnamefont {D.~C.}\ \bibnamefont {Glattli}}, \bibinfo
  {author} {\bibfnamefont {B.}~\bibnamefont {Pla{\c{c}}ais}}, \bibinfo {author}
  {\bibfnamefont {G.}~\bibnamefont {F{\` e}ve}}, \bibinfo {author}
  {\bibfnamefont {M.}~\bibnamefont {Albert}}, \bibinfo {author} {\bibfnamefont
  {C.}~\bibnamefont {Flindt}}, \ and\ \bibinfo {author} {\bibfnamefont
  {M.}~\bibnamefont {B{\"u}ttiker}},\ }\href@noop {} {\bibfield  {journal}
  {\bibinfo  {journal} {Phys. Rev. B}\ }\textbf {\bibinfo {volume} {85}},\
  \bibinfo {pages} {165438} (\bibinfo {year} {2012})}\BibitemShut {NoStop}%
\bibitem [{\citenamefont {Sherkunov}\ \emph {et~al.}(2012)\citenamefont
  {Sherkunov}, \citenamefont {{d'Ambrumenil}}, \citenamefont {Samuelsson},\
  and\ \citenamefont {B{\"u}ttiker}}]{sherkunov_optimal_2012}%
  \BibitemOpen
  \bibfield  {author} {\bibinfo {author} {\bibfnamefont {Y.}~\bibnamefont
  {Sherkunov}}, \bibinfo {author} {\bibfnamefont {N.}~\bibnamefont
  {{d'Ambrumenil}}}, \bibinfo {author} {\bibfnamefont {P.}~\bibnamefont
  {Samuelsson}}, \ and\ \bibinfo {author} {\bibfnamefont {M.}~\bibnamefont
  {B{\"u}ttiker}},\ }\href@noop {} {\bibfield  {journal} {\bibinfo  {journal}
  {Phys. Rev. B}\ }\textbf {\bibinfo {volume} {85}},\ \bibinfo {pages}
  {081108(R)} (\bibinfo {year} {2012})}\BibitemShut {NoStop}%
\bibitem [{\citenamefont {Gabelli}\ and\ \citenamefont
  {Reulet}(2012)}]{gabelli_shaping_2012}%
  \BibitemOpen
  \bibfield  {author} {\bibinfo {author} {\bibfnamefont {J.}~\bibnamefont
  {Gabelli}}\ and\ \bibinfo {author} {\bibfnamefont {B.}~\bibnamefont
  {Reulet}},\ }\href@noop {} {\bibfield  {journal} {\bibinfo  {journal}
  {{arXiv:1205.3638}}\ } (\bibinfo {year} {2012})}\BibitemShut {NoStop}%
\bibitem [{\citenamefont {Vanevi\'{c}}\ \emph {et~al.}(2007)\citenamefont
  {Vanevi\'{c}}, \citenamefont {{Y}u. V.~Nazarov},\ and\ \citenamefont
  {Belzig}}]{VanevicNazarovBelzigPRL99-07}%
  \BibitemOpen
  \bibfield  {author} {\bibinfo {author} {\bibfnamefont {M.}~\bibnamefont
  {Vanevi\'{c}}}, \bibinfo {author} {\bibnamefont {{Y}u. V.~Nazarov}}, \ and\
  \bibinfo {author} {\bibfnamefont {W.}~\bibnamefont {Belzig}},\ }\href@noop {}
  {\bibfield  {journal} {\bibinfo  {journal} {Phys. Rev. Lett.}\ }\textbf
  {\bibinfo {volume} {99}},\ \bibinfo {pages} {076601} (\bibinfo {year}
  {2007})}\BibitemShut {NoStop}%
\bibitem [{\citenamefont {Vanevi\'{c}}\ \emph {et~al.}(2008)\citenamefont
  {Vanevi\'{c}}, \citenamefont {{Y}u. V.~Nazarov},\ and\ \citenamefont
  {Belzig}}]{VanevicNazarovBelzigPRB78-08}%
  \BibitemOpen
  \bibfield  {author} {\bibinfo {author} {\bibfnamefont {M.}~\bibnamefont
  {Vanevi\'{c}}}, \bibinfo {author} {\bibnamefont {{Y}u. V.~Nazarov}}, \ and\
  \bibinfo {author} {\bibfnamefont {W.}~\bibnamefont {Belzig}},\ }\href@noop {}
  {\bibfield  {journal} {\bibinfo  {journal} {Phys. Rev. B}\ }\textbf {\bibinfo
  {volume} {78}},\ \bibinfo {pages} {245308} (\bibinfo {year}
  {2008})}\BibitemShut {NoStop}%
\bibitem [{\citenamefont {Beenakker}\ \emph {et~al.}(2003)\citenamefont
  {Beenakker}, \citenamefont {Emary}, \citenamefont {Kindermann},\ and\
  \citenamefont {van Velsen}}]{beenakker_proposal_2003}%
  \BibitemOpen
  \bibfield  {author} {\bibinfo {author} {\bibfnamefont {C.~W.~J.}\
  \bibnamefont {Beenakker}}, \bibinfo {author} {\bibfnamefont {C.}~\bibnamefont
  {Emary}}, \bibinfo {author} {\bibfnamefont {M.}~\bibnamefont {Kindermann}}, \
  and\ \bibinfo {author} {\bibfnamefont {J.~L.}\ \bibnamefont {van Velsen}},\
  }\href@noop {} {\bibfield  {journal} {\bibinfo  {journal} {Phys. Rev. Lett.}\
  }\textbf {\bibinfo {volume} {91}},\ \bibinfo {pages} {147901} (\bibinfo
  {year} {2003})}\BibitemShut {NoStop}%
\bibitem [{\citenamefont {Samuelsson}\ \emph {et~al.}(2004)\citenamefont
  {Samuelsson}, \citenamefont {Sukhorukov},\ and\ \citenamefont
  {B{\"u}ttiker}}]{samuelsson_two-particle_2004}%
  \BibitemOpen
  \bibfield  {author} {\bibinfo {author} {\bibfnamefont {P.}~\bibnamefont
  {Samuelsson}}, \bibinfo {author} {\bibfnamefont {E.~V.}\ \bibnamefont
  {Sukhorukov}}, \ and\ \bibinfo {author} {\bibfnamefont {M.}~\bibnamefont
  {B{\"u}ttiker}},\ }\href@noop {} {\bibfield  {journal} {\bibinfo  {journal}
  {Phys. Rev. Lett.}\ }\textbf {\bibinfo {volume} {92}},\ \bibinfo {pages}
  {026805} (\bibinfo {year} {2004})}\BibitemShut {NoStop}%
\bibitem [{\citenamefont {Keeling}\ \emph {et~al.}(2008)\citenamefont
  {Keeling}, \citenamefont {Shytov},\ and\ \citenamefont
  {Levitov}}]{KeelingShytovLevitovPRL101-08}%
  \BibitemOpen
  \bibfield  {author} {\bibinfo {author} {\bibfnamefont {J.}~\bibnamefont
  {Keeling}}, \bibinfo {author} {\bibfnamefont {A.~V.}\ \bibnamefont {Shytov}},
  \ and\ \bibinfo {author} {\bibfnamefont {L.~S.}\ \bibnamefont {Levitov}},\
  }\href@noop {} {\bibfield  {journal} {\bibinfo  {journal} {Phys. Rev. Lett.}\
  }\textbf {\bibinfo {volume} {101}},\ \bibinfo {pages} {196404} (\bibinfo
  {year} {2008})}\BibitemShut {NoStop}%
\bibitem [{\citenamefont {Keeling}\ \emph {et~al.}(2006)\citenamefont
  {Keeling}, \citenamefont {Klich},\ and\ \citenamefont
  {Levitov}}]{KeelingKlichLevitovPRL97-06}%
  \BibitemOpen
  \bibfield  {author} {\bibinfo {author} {\bibfnamefont {J.}~\bibnamefont
  {Keeling}}, \bibinfo {author} {\bibfnamefont {I.}~\bibnamefont {Klich}}, \
  and\ \bibinfo {author} {\bibfnamefont {L.~S.}\ \bibnamefont {Levitov}},\
  }\href@noop {} {\bibfield  {journal} {\bibinfo  {journal} {Phys. Rev. Lett.}\
  }\textbf {\bibinfo {volume} {97}},\ \bibinfo {pages} {116403} (\bibinfo
  {year} {2006})}\BibitemShut {NoStop}%
\bibitem [{\citenamefont {Ivanov}\ \emph {et~al.}(1997)\citenamefont {Ivanov},
  \citenamefont {Lee},\ and\ \citenamefont
  {Levitov}}]{IvanovLeeLevitovPRB56-97}%
  \BibitemOpen
  \bibfield  {author} {\bibinfo {author} {\bibfnamefont {D.~A.}\ \bibnamefont
  {Ivanov}}, \bibinfo {author} {\bibfnamefont {H.~W.}\ \bibnamefont {Lee}}, \
  and\ \bibinfo {author} {\bibfnamefont {L.~S.}\ \bibnamefont {Levitov}},\
  }\href@noop {} {\bibfield  {journal} {\bibinfo  {journal} {Phys. Rev. B}\
  }\textbf {\bibinfo {volume} {56}},\ \bibinfo {pages} {6839} (\bibinfo {year}
  {1997})}\BibitemShut {NoStop}%
\bibitem [{\citenamefont {Lesovik}\ and\ \citenamefont
  {Levitov}(1994)}]{LesovikLevitovPRL72-94}%
  \BibitemOpen
  \bibfield  {author} {\bibinfo {author} {\bibfnamefont {G.~B.}\ \bibnamefont
  {Lesovik}}\ and\ \bibinfo {author} {\bibfnamefont {L.~S.}\ \bibnamefont
  {Levitov}},\ }\href@noop {} {\bibfield  {journal} {\bibinfo  {journal} {Phys.
  Rev. Lett.}\ }\textbf {\bibinfo {volume} {72}},\ \bibinfo {pages} {538}
  (\bibinfo {year} {1994})}\BibitemShut {NoStop}%
\bibitem [{\citenamefont {Pedersen}\ and\ \citenamefont
  {B{\"u}ttiker}(1998)}]{PedersenButtikerPRB58-98}%
  \BibitemOpen
  \bibfield  {author} {\bibinfo {author} {\bibfnamefont {M.~H.}\ \bibnamefont
  {Pedersen}}\ and\ \bibinfo {author} {\bibfnamefont {M.}~\bibnamefont
  {B{\"u}ttiker}},\ }\href@noop {} {\bibfield  {journal} {\bibinfo  {journal}
  {Phys. Rev. B}\ }\textbf {\bibinfo {volume} {58}},\ \bibinfo {pages} {12993}
  (\bibinfo {year} {1998})}\BibitemShut {NoStop}%
\bibitem [{\citenamefont {Schoelkopf}\ \emph {et~al.}(1998)\citenamefont
  {Schoelkopf}, \citenamefont {Kozhevnikov}, \citenamefont {Prober},\ and\
  \citenamefont {Rooks}}]{SchoelkopfKozhevnikovProberRooksPRL80-98}%
  \BibitemOpen
  \bibfield  {author} {\bibinfo {author} {\bibfnamefont {R.~J.}\ \bibnamefont
  {Schoelkopf}}, \bibinfo {author} {\bibfnamefont {A.~A.}\ \bibnamefont
  {Kozhevnikov}}, \bibinfo {author} {\bibfnamefont {D.~E.}\ \bibnamefont
  {Prober}}, \ and\ \bibinfo {author} {\bibfnamefont {M.~J.}\ \bibnamefont
  {Rooks}},\ }\href@noop {} {\bibfield  {journal} {\bibinfo  {journal} {Phys.
  Rev. Lett.}\ }\textbf {\bibinfo {volume} {80}},\ \bibinfo {pages} {2437}
  (\bibinfo {year} {1998})}\BibitemShut {NoStop}%
\bibitem [{\citenamefont {Kozhevnikov}\ \emph {et~al.}(2000)\citenamefont
  {Kozhevnikov}, \citenamefont {Schoelkopf},\ and\ \citenamefont
  {Prober}}]{KozhevnikovProberPRL84-99}%
  \BibitemOpen
  \bibfield  {author} {\bibinfo {author} {\bibfnamefont {A.~A.}\ \bibnamefont
  {Kozhevnikov}}, \bibinfo {author} {\bibfnamefont {R.~J.}\ \bibnamefont
  {Schoelkopf}}, \ and\ \bibinfo {author} {\bibfnamefont {D.~E.}\ \bibnamefont
  {Prober}},\ }\href@noop {} {\bibfield  {journal} {\bibinfo  {journal} {Phys.
  Rev. Lett.}\ }\textbf {\bibinfo {volume} {84}},\ \bibinfo {pages} {3398}
  (\bibinfo {year} {2000})}\BibitemShut {NoStop}%
\bibitem [{\citenamefont {Reydellet}\ \emph {et~al.}(2003)\citenamefont
  {Reydellet}, \citenamefont {Roche}, \citenamefont {Glattli}, \citenamefont
  {Etienne},\ and\ \citenamefont
  {Jin}}]{ReydelletRocheGlattliEtienneJinPRL90-03}%
  \BibitemOpen
  \bibfield  {author} {\bibinfo {author} {\bibfnamefont {L.-H.}\ \bibnamefont
  {Reydellet}}, \bibinfo {author} {\bibfnamefont {P.}~\bibnamefont {Roche}},
  \bibinfo {author} {\bibfnamefont {D.~C.}\ \bibnamefont {Glattli}}, \bibinfo
  {author} {\bibfnamefont {B.}~\bibnamefont {Etienne}}, \ and\ \bibinfo
  {author} {\bibfnamefont {Y.}~\bibnamefont {Jin}},\ }\href@noop {} {\bibfield
  {journal} {\bibinfo  {journal} {Phys. Rev. Lett.}\ }\textbf {\bibinfo
  {volume} {90}},\ \bibinfo {pages} {176803} (\bibinfo {year}
  {2003})}\BibitemShut {NoStop}%
\bibitem [{\citenamefont {Rychkov}\ \emph {et~al.}(2005)\citenamefont
  {Rychkov}, \citenamefont {Polianski},\ and\ \citenamefont
  {B{\"u}ttiker}}]{RychkovPolianskiButtikerPRB72-05}%
  \BibitemOpen
  \bibfield  {author} {\bibinfo {author} {\bibfnamefont {V.~S.}\ \bibnamefont
  {Rychkov}}, \bibinfo {author} {\bibfnamefont {M.~L.}\ \bibnamefont
  {Polianski}}, \ and\ \bibinfo {author} {\bibfnamefont {M.}~\bibnamefont
  {B{\"u}ttiker}},\ }\href@noop {} {\bibfield  {journal} {\bibinfo  {journal}
  {Phys. Rev. B}\ }\textbf {\bibinfo {volume} {72}},\ \bibinfo {pages} {155326}
  (\bibinfo {year} {2005})}\BibitemShut {NoStop}%
\bibitem [{\citenamefont {Polianski}\ \emph {et~al.}(2005)\citenamefont
  {Polianski}, \citenamefont {Samuelsson},\ and\ \citenamefont
  {B{\"u}ttiker}}]{PolianskiSamuelssonBuettikerPRB72-05}%
  \BibitemOpen
  \bibfield  {author} {\bibinfo {author} {\bibfnamefont {M.~L.}\ \bibnamefont
  {Polianski}}, \bibinfo {author} {\bibfnamefont {P.}~\bibnamefont
  {Samuelsson}}, \ and\ \bibinfo {author} {\bibfnamefont {M.}~\bibnamefont
  {B{\"u}ttiker}},\ }\href@noop {} {\bibfield  {journal} {\bibinfo  {journal}
  {Phys. Rev. B}\ }\textbf {\bibinfo {volume} {72}},\ \bibinfo {pages}
  {R161302} (\bibinfo {year} {2005})}\BibitemShut {NoStop}%
\bibitem [{\citenamefont {Samuelsson}\ and\ \citenamefont
  {B{\"u}ttiker}(2005)}]{samuelsson_dynamic_2005}%
  \BibitemOpen
  \bibfield  {author} {\bibinfo {author} {\bibfnamefont {P.}~\bibnamefont
  {Samuelsson}}\ and\ \bibinfo {author} {\bibfnamefont {M.}~\bibnamefont
  {B{\"u}ttiker}},\ }\href@noop {} {\bibfield  {journal} {\bibinfo  {journal}
  {Phys. Rev. B}\ }\textbf {\bibinfo {volume} {71}},\ \bibinfo {pages} {245317}
  (\bibinfo {year} {2005})}\BibitemShut {NoStop}%
\bibitem [{\citenamefont {Beenakker}\ \emph {et~al.}(2005)\citenamefont
  {Beenakker}, \citenamefont {Titov},\ and\ \citenamefont
  {Trauzettel}}]{beenakker_optimal_2005}%
  \BibitemOpen
  \bibfield  {author} {\bibinfo {author} {\bibfnamefont {C.~W.~J.}\
  \bibnamefont {Beenakker}}, \bibinfo {author} {\bibfnamefont {M.}~\bibnamefont
  {Titov}}, \ and\ \bibinfo {author} {\bibfnamefont {B.}~\bibnamefont
  {Trauzettel}},\ }\href@noop {} {\bibfield  {journal} {\bibinfo  {journal}
  {Phys. Rev. Lett.}\ }\textbf {\bibinfo {volume} {94}},\ \bibinfo {pages}
  {186804} (\bibinfo {year} {2005})}\BibitemShut {NoStop}%
\bibitem [{\citenamefont {Lebedev}\ \emph {et~al.}(2005)\citenamefont
  {Lebedev}, \citenamefont {Lesovik},\ and\ \citenamefont
  {Blatter}}]{LebedevLesovikBlatterPRB72-05}%
  \BibitemOpen
  \bibfield  {author} {\bibinfo {author} {\bibfnamefont {A.~V.}\ \bibnamefont
  {Lebedev}}, \bibinfo {author} {\bibfnamefont {G.~B.}\ \bibnamefont
  {Lesovik}}, \ and\ \bibinfo {author} {\bibfnamefont {G.}~\bibnamefont
  {Blatter}},\ }\href@noop {} {\bibfield  {journal} {\bibinfo  {journal} {Phys.
  Rev. B}\ }\textbf {\bibinfo {volume} {72}},\ \bibinfo {pages} {245314}
  (\bibinfo {year} {2005})}\BibitemShut {NoStop}%
\bibitem [{\citenamefont {Splettstoesser}\ \emph {et~al.}(2009)\citenamefont
  {Splettstoesser}, \citenamefont {Moskalets},\ and\ \citenamefont
  {B{\"u}ttiker}}]{splettstoesser_two-particle_2009}%
  \BibitemOpen
  \bibfield  {author} {\bibinfo {author} {\bibfnamefont {J.}~\bibnamefont
  {Splettstoesser}}, \bibinfo {author} {\bibfnamefont {M.}~\bibnamefont
  {Moskalets}}, \ and\ \bibinfo {author} {\bibfnamefont {M.}~\bibnamefont
  {B{\"u}ttiker}},\ }\href@noop {} {\bibfield  {journal} {\bibinfo  {journal}
  {Phys. Rev. Lett.}\ }\textbf {\bibinfo {volume} {103}},\ \bibinfo {pages}
  {076804} (\bibinfo {year} {2009})}\BibitemShut {NoStop}%
\bibitem [{\citenamefont {Moskalets}\ and\ \citenamefont
  {B{\"u}ttiker}(2011)}]{moskalets_spectroscopy_2011}%
  \BibitemOpen
  \bibfield  {author} {\bibinfo {author} {\bibfnamefont {M.}~\bibnamefont
  {Moskalets}}\ and\ \bibinfo {author} {\bibfnamefont {M.}~\bibnamefont
  {B{\"u}ttiker}},\ }\href@noop {} {\bibfield  {journal} {\bibinfo  {journal}
  {Phys. Rev. B}\ }\textbf {\bibinfo {volume} {83}},\ \bibinfo {pages} {035316}
  (\bibinfo {year} {2011})}\BibitemShut {NoStop}%
\bibitem [{\citenamefont {Grenier}\ \emph {et~al.}(2011)\citenamefont
  {Grenier}, \citenamefont {Herv{\'e}}, \citenamefont {Bocquillon},
  \citenamefont {Parmentier}, \citenamefont {Pla{\c{c}}ais}, \citenamefont
  {Berroir}, \citenamefont {F{\`e}ve},\ and\ \citenamefont
  {Degiovanni}}]{grenier_single-electron_2011}%
  \BibitemOpen
  \bibfield  {author} {\bibinfo {author} {\bibfnamefont {C.}~\bibnamefont
  {Grenier}}, \bibinfo {author} {\bibfnamefont {R.}~\bibnamefont {Herv{\'e}}},
  \bibinfo {author} {\bibfnamefont {E.}~\bibnamefont {Bocquillon}}, \bibinfo
  {author} {\bibfnamefont {F.~D.}\ \bibnamefont {Parmentier}}, \bibinfo
  {author} {\bibfnamefont {B.}~\bibnamefont {Pla{\c{c}}ais}}, \bibinfo {author}
  {\bibfnamefont {J.~M.}\ \bibnamefont {Berroir}}, \bibinfo {author}
  {\bibfnamefont {G.}~\bibnamefont {F{\`e}ve}}, \ and\ \bibinfo {author}
  {\bibfnamefont {P.}~\bibnamefont {Degiovanni}},\ }\href@noop {} {\bibfield
  {journal} {\bibinfo  {journal} {New J. Phys.}\ }\textbf {\bibinfo {volume}
  {13}},\ \bibinfo {pages} {093007} (\bibinfo {year} {2011})}\BibitemShut
  {NoStop}%
\end{thebibliography}%

\end{document}